\documentclass[12pt,twoside,fleqn]{article}
\usepackage{epsfig}
\usepackage{amssymb,espcrc1}
\newdimen\htxtw
\htxtw=\textwidth
\divide\htxtw by 2

\vspace{2cm}
\title{Towards the Quark-Gluon-Plasma }

\author{Johanna Stachel \\
Physikalisches Institut der Universit\"at Heidelberg, 69120 Heidelberg,
Germany  
 }

\begin{document}
\maketitle

\begin{abstract}

\end{abstract}
  
\section{INTRODUCTION}
\enlargethispage{\baselineskip}
A new phase of hadronic matter is expected to form at high temperature
and/or high baryon density. This is based on considerations in
perturbative quantum chromo dynamics (QCD) and in chiral perturbation
theory. The most quantitative information about  where this transition is
expected to happen comes from numerical solutions of QCD on the lattice
where large progress was made during the past few years
\cite{lattice}. The systematic influence of the size of the space-time
lattice, of the quark masses, and of the way to implement fermions on
the lattice has been explored. Extrapolations to the chiral limit of a
vanishing mass of the pseudoscalar Goldstone boson, the pion, give a
critical temperature of 20 \% of the $\rho$ mass, i.e. 154 MeV
\cite{lattice1,lattice2}. The calculations indicate that at this
temperature deconfinement is lifted and the quark condensate disappears
leading to the restoration of chiral symmetry. The quark-gluon plasma
(QGP) is formed. This is associated with a rapid rise in the number of
degrees of freedom, and therefore the energy and entropy density, by
typically one order of magnitude over a temperature interval of the
order of 10 MeV.

This quark-gluon plasma presumably existed in the early universe until
it hadronized at an age of ten microseconds and it may exist today in
the interior of neutron stars.  In order to produce this new phase in
experiments in the laboratory collisions of heavy nuclei at the highest
possible energies are considered the most likely tool. An experimental
program started at the Brookhaven AGS and at the CERN SPS in 1986  and
since 1992 and 1994, respectively, collisions of the heaviest nuclei are
studied.

For heating and population of new degrees of freedom in the nuclear
fireball the relevant number is the energy that is available in the
c.m. system (without counting the masses of the constituent
nucleons). For Au+Au collisions at the AGS and for Pb+Pb collisions at
the SPS the respective numbers are 600 and 3200 GeV per collision. After
the fireball hadronizes these numbers materialize in terms of about 900
and 2400 hadrons in the final state. It is interesting to note that the
two accelerators give access to two quite different regimes in the final
state: at the AGS the number of nucleons and pions is about comparable
while at the SPS pions dominate by a factor of six. In either case, the
nucleons are dominantly the constituent nucleons of projectile and
target.


In the following the experimental results available after one decade of
studying ultra relativistic heavy ion collisions will be reviewed and
discussed vis-a-vis the expected phase transition to the QGP.

\section{EQUILIBRATION}

One of the key questions is the issue whether the fireball ever reaches
equilibrium during its evolution. For the hadronic final state this
question can be addressed in terms of asking whether the hadron yields
are consistent with expectations for a statistical ensemble. 

\begin{figure}[b]

\vspace{-1.6cm}

\hspace*{2.cm}
\epsfig{file=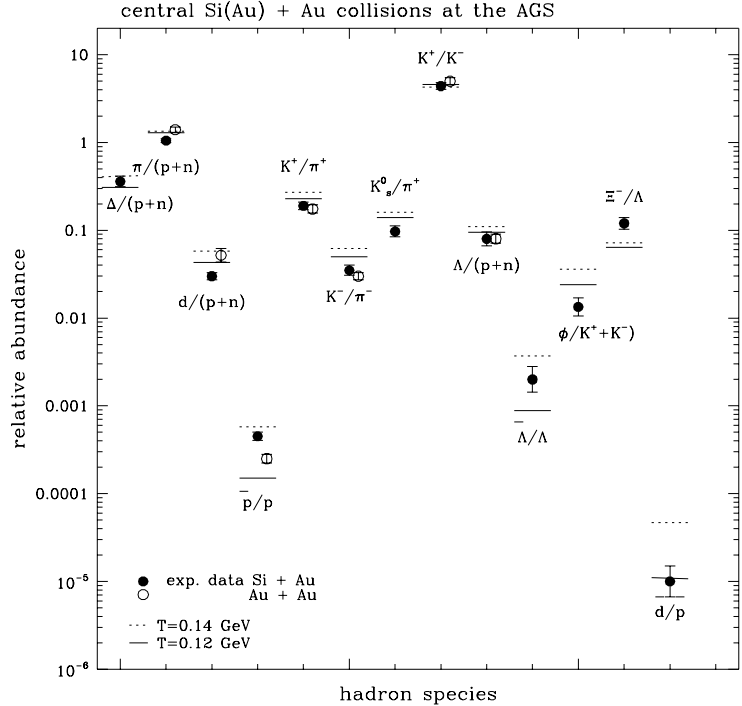,width=12cm}

\vspace{-1.8cm}

\caption{\small Hadron yields for central 14.6 A GeV/c Si+Au and 11 A
GeV/c Au+Au collisions at
the AGS compared to a thermal model calculation for two
temperatures. Figure from 
\protect\cite{qm96}.
} 
\label{figyags}
\end{figure}

In Figure~\ref{figyags} is shown the systematics of relative abundances
of produced hadrons from the AGS after integrating over longitudinal and
transverse momenta as much as the data allow. In this way the analysis
becomes independent of dynamic effects such as expansion. The relative
abundances span about 9 orders of magnitude comparing pion and nucleon
yields to those of antideuterons. The main difference to results from
nucleon nucleon collisions is a significant enhancement of strangeness,
e.g. of a factor two to three in the kaon yield relative to pions in
central Au+Au collisions. Central Si+Au and Au+Au collisions lead to
very similar results.

The experimental data are compared to calculations \cite{thermal1} for a
grand canonical ensemble and there are two free parameters adjusted to
the data: a temperature T which is found to be between 120 and 140 MeV
and a baryochemical potential $\mu_b$ = 540 MeV driven by the pion to
nucleon ratio. One can see that the data are rather well described by
this simple assumption.

\begin{figure}

\vspace{-1.6cm}

\hspace*{1.5cm}
\epsfig{file=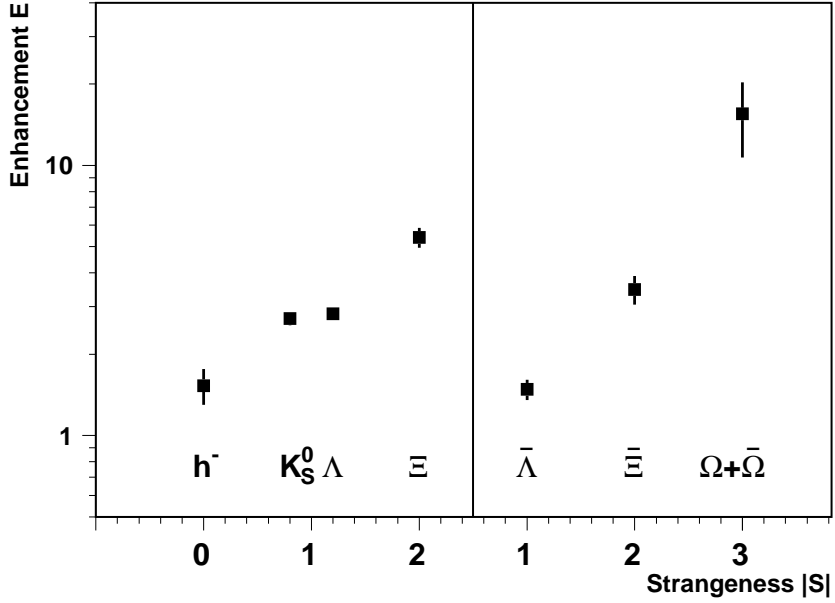,width=12cm}

\vspace{-.9cm}

\caption{\small Enhancement of strange particle production in
mid-central to central 158 A GeV/c Pb+Pb collisions relative to
p+Au collisions. Figure from
\protect\cite{wa97}.
} 
\label{figwa97}
\end{figure}

This observation of enhanced strangeness production becomes even more
dramatic in central Pb+Pb collisions at the SPS where also multiply
strange baryons have been studied. Figure~\ref{figwa97} shows for
semicentral to central Pb+Pb collisions an enhancement for hadrons
carrying strangeness rising dramatically with increasing strangeness.
Here the enhancement is defined by scaling the yield of various produced
particles to the number of nucleons participating in the collision, as
defined by the impact parameter, and normalizing the result for p+Pb
collisions to unity. The triply strange omega baryons are produced about
a factor 15 more frequently than expected for a simple linear scaling
with the number of participating nucleons.

\enlargethispage{-\baselineskip}

\begin{figure}[b]

\vspace{-1.6cm}

\hspace*{0.5cm}
\epsfig{file=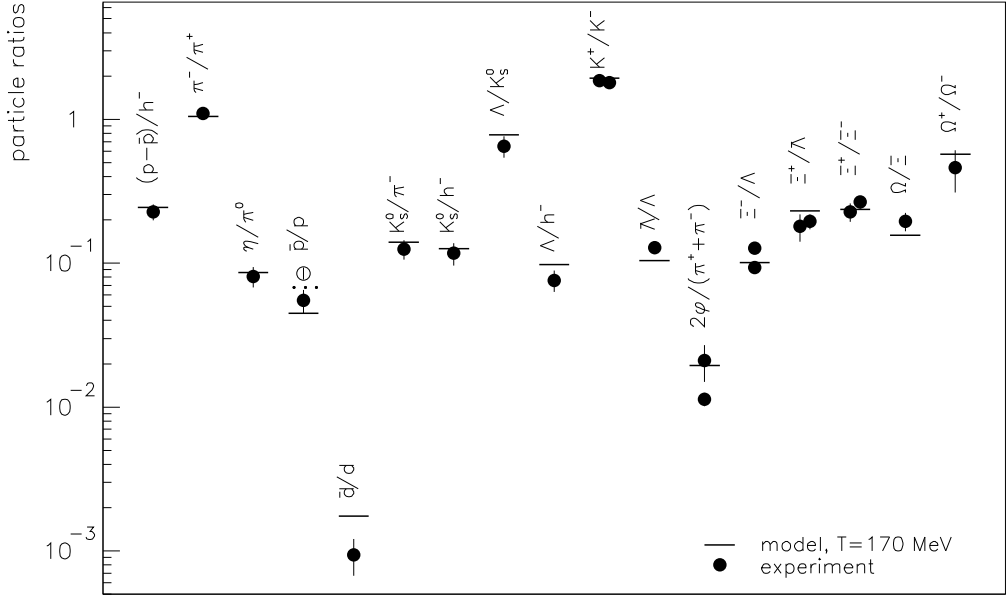,width=15cm}

\vspace{-1.8cm}

\caption{\small Hadron yields for central Pb+Pb collisions at 158 A
GeV/c at the SPS compared to a fit with a thermal model; Figure from 
\protect\cite{therm_pbpb}.
} 
\label{figysps}
\end{figure}

A summary of the yields of produced particles for central Pb+Pb
collisions at the SPS is shown in Figure~\ref{figysps}. This combines
data from all heavy ion experiments. Again the experimental data are
compared to a calculation for a grand canonical ensemble, including the
know hadron spectrum up to 1.5 and 2 GeV/c$^2$ for mesons and baryons.
And again the agreement is very good. The resulting two fit parameters
are T = 170 MeV and $\mu_b$ = 280 MeV. The higher temperature and lower
baryochemical potential as compared to the AGS fit are reflected in a
much smaller dynamic range of yields of produced particles spanning in
this case only about 4 orders of magnitude between pion and
antideuteron. There is no indication that strangeness is suppressed
vis-a-vis the equilibrium assumption and this is in marked contrast to
nucleon-nucleon collisions where a similar analysis is only successful
if one allows for explicit suppression by about a factor of two of
particles carrying strangeness \cite{becattini}.

\begin{figure}[tb]


\hspace*{1.cm}
\epsfig{file=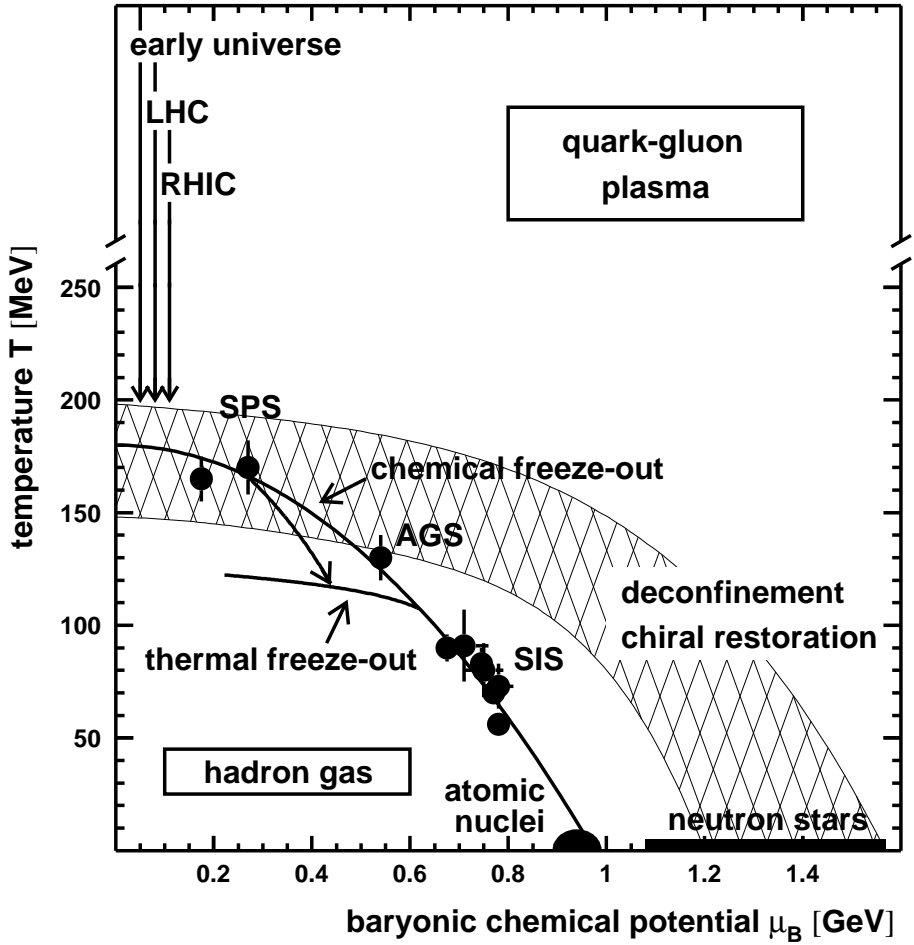,width=12cm,bbllx=38,bblly=150,bburx=555,bbury=687}


\caption{\small Phase diagram of hadronic matter and hadrochemical
freeze-out points for heavy ion collisions at SIS, AGS and SPS. The
hatched region indicates the expectation
\protect\cite{phase_gerry}
for the phase boundary based on lattice QCD calculations at
$\mu_b$=0. The arrow from chemical to thermal freeze-out curve for the
SPS is calculated for expansion with constant entropy per baryon. Figure from  
\protect\cite{qm97,averbeck}.} 
\label{figphased}
\end{figure}

In summary, one can conclude that hadron yields at AGS and SPS are
consistent with the assumption of equilibration in terms of produced
particle species.  The values of temperature and baryochemical potential
resulting from the fit with a statistical ensemble thus determine the
points in the phase diagram when, during the evolution of the fireball,
the hadron yields are frozen. This happens when inelastic collisions
cease and this point is termed 'chemical freeze-out'. In order to put
these freeze-out points into perspective relative to the expected phase
transition, the result for the critical point from lattice QCD has to be
extended towards finite baryon density. This can be done by constructing
the phase boundary between a complete hadron gas as used e.g. in the
yield calculations and an ideal gas of quarks and gluons using the
Gibbs' conditions. Such a calculation \cite{phase_gerry} is shown in
Figure~\ref{figphased} as the hatched band in comparison to the
freeze-out points determined from yield ratios. There, the lower boundary
of the hatched band reproduces the currently best result from lattice
QCD at zero baryon density, the upper edge reflects the maximum
systematic error still possibly contained in these calculations \cite{laer}.

One notices that the chemical freeze-out points at the AGS and SPS are
near or at the phase boundary while the points from the lower SIS energy
show that there the system is far from the critical curve. It was noted
recently \cite{cley} that the empirical curve outlined by the
experimental points corresponds to a constant energy per hadron of about
1 GeV.

\begin{figure}[b]

\vspace{-0.6cm}

\hspace*{2.cm}
\epsfig{file=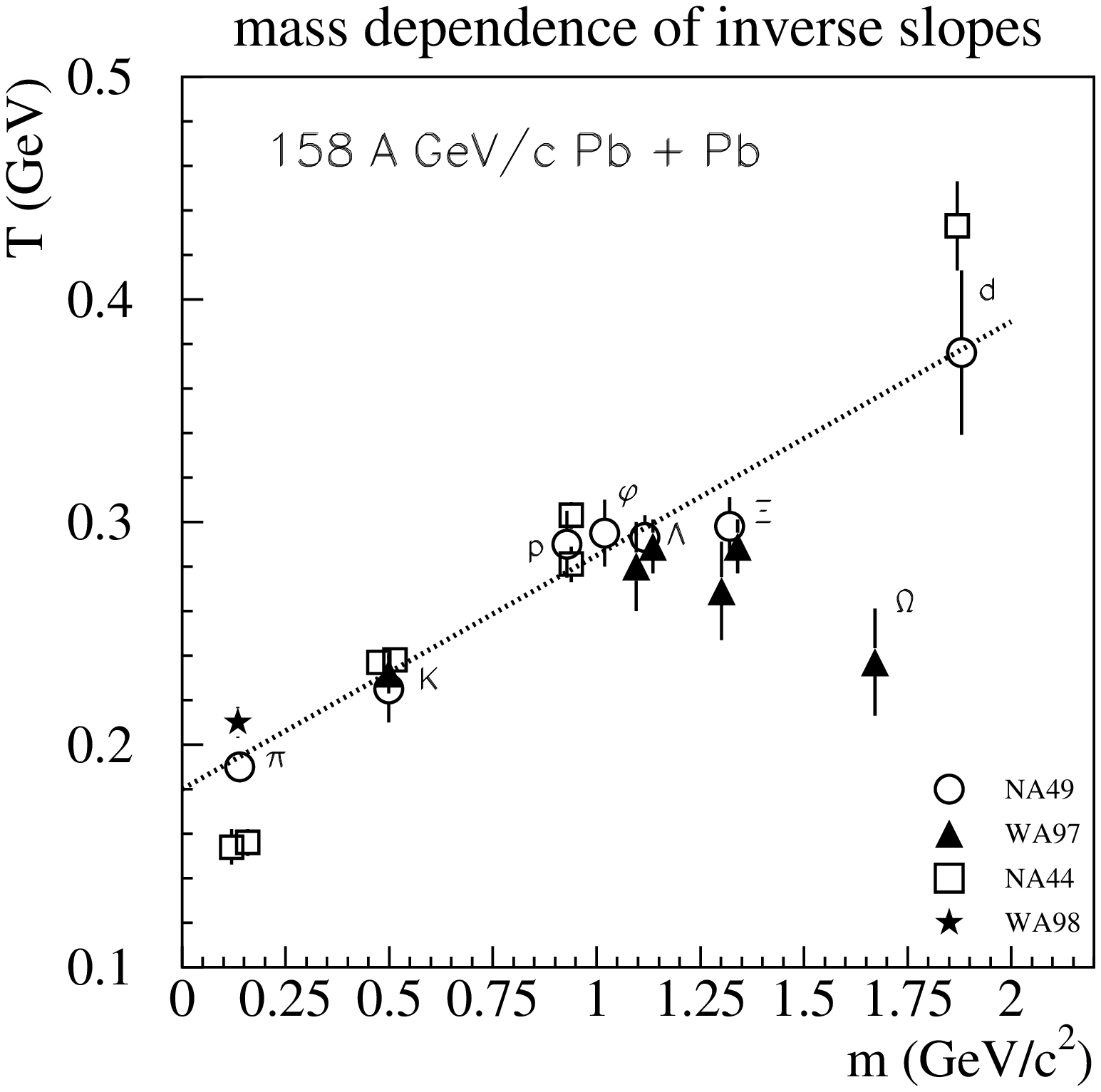,width=12cm}

\vspace{-2.5cm}

\caption{\small Systematics of inverse slope constants fitted to hadron
spectra from central Pb+Pb collisions at
the SPS; experimental data from
\protect\cite{na44sl,na49sl,wa97sl,wa98sl}.
 }
\label{figslop}
\end{figure}

The question of 'thermal freeze-out' or the point when  also elastic collisions seize can be addressed by inspecting the hadron
spectra. Again, there is a distinct difference to the behavior know from
p+p collisions. Both at the AGS and SPS slopes of distributions in
transverse mass m$_{\rm t}$ = $\sqrt{{\rm p_t^2 + m^2}}$ rise
significantly with increasing mass. This is shown for the case of
central Pb+Pb collisions at the SPS in Figure~\ref{figslop} combining
data from the different experiments. The dotted line indicates the
approximate behavior.

Such an ordering in mass arises when the random thermal motion is
superimposed to e.g. a common transverse expansion of the system
\cite{sr} and such a picture has already successfully accounted for the
spectra from Si and S induced reactions at the AGS and SPS
\cite{thermal1,thermal2}. However it was noted that a range of
combinations of temperatures and appropriate expansion velocities can
account for the experimental data \cite{esumi}. This ambiguity can be
avoided by choosing a certain initial condition and an equation of
state; then for every instant in the systems evolution a unique
combination of temperature and velocity (profile) is given. Using this
approach it was shown \cite{qm97} that a temperature of 160 MeV cannot
describe the Pb+Pb SPS hadron spectra. Rather a temperature of 120 MeV
is required. The same temperature and velocity profile give a good
description of the AGS Au+Au hadron spectra both in terms of overall
slope and shape.

One clear exception from the simple scaling of spectral slopes with mass
is visible in Fig.~\ref{figslop}: the slope of the $\Omega$ spectrum is
significantly lower than expected from the systematic trend although it
is also much above a thermal distribution with T = 120 MeV. This could
be connected possibly with a smaller elastic cross section for the
$\Omega$ or could conceivably also point to a different spatial
distribution of the $\Omega$'s in the fireball (at the inside the
velocities are smaller).

Information on the expansion of the fireball can also be gained from the
two-particle correlations. The data can be sensitive both to the final
volume as well as to the expansion dynamics. 

\begin{figure}[t]
  \vspace{-.7cm}  
  \begin{flushleft}
    \begin{minipage}{0.5\linewidth}
      \epsfig{figure=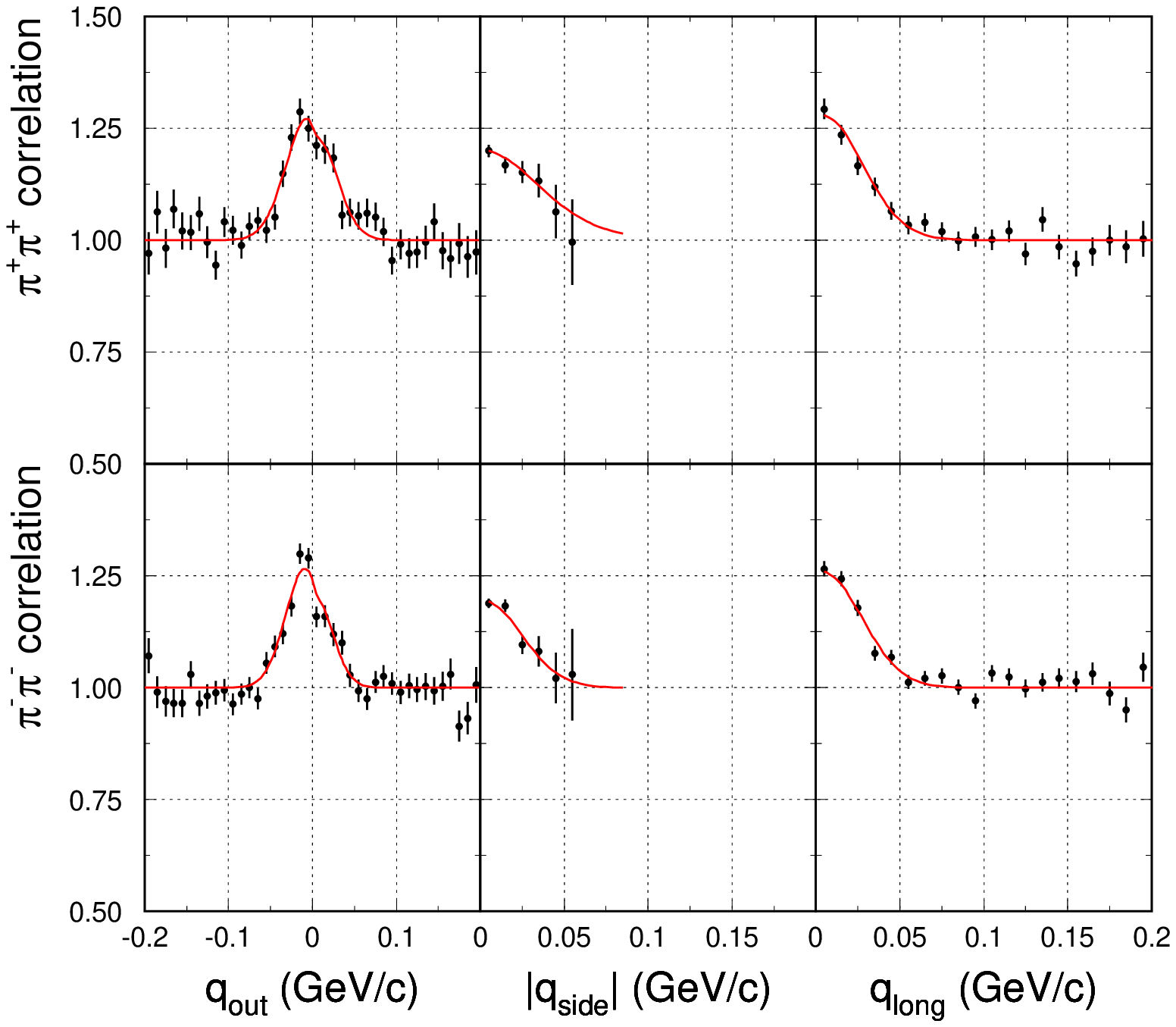,width=1.\linewidth}
    \end{minipage}
  \end{flushleft}
  \vspace{-9.2cm}
  \begin{flushright}
    \begin{minipage}{0.5\linewidth}
      \epsfig{figure=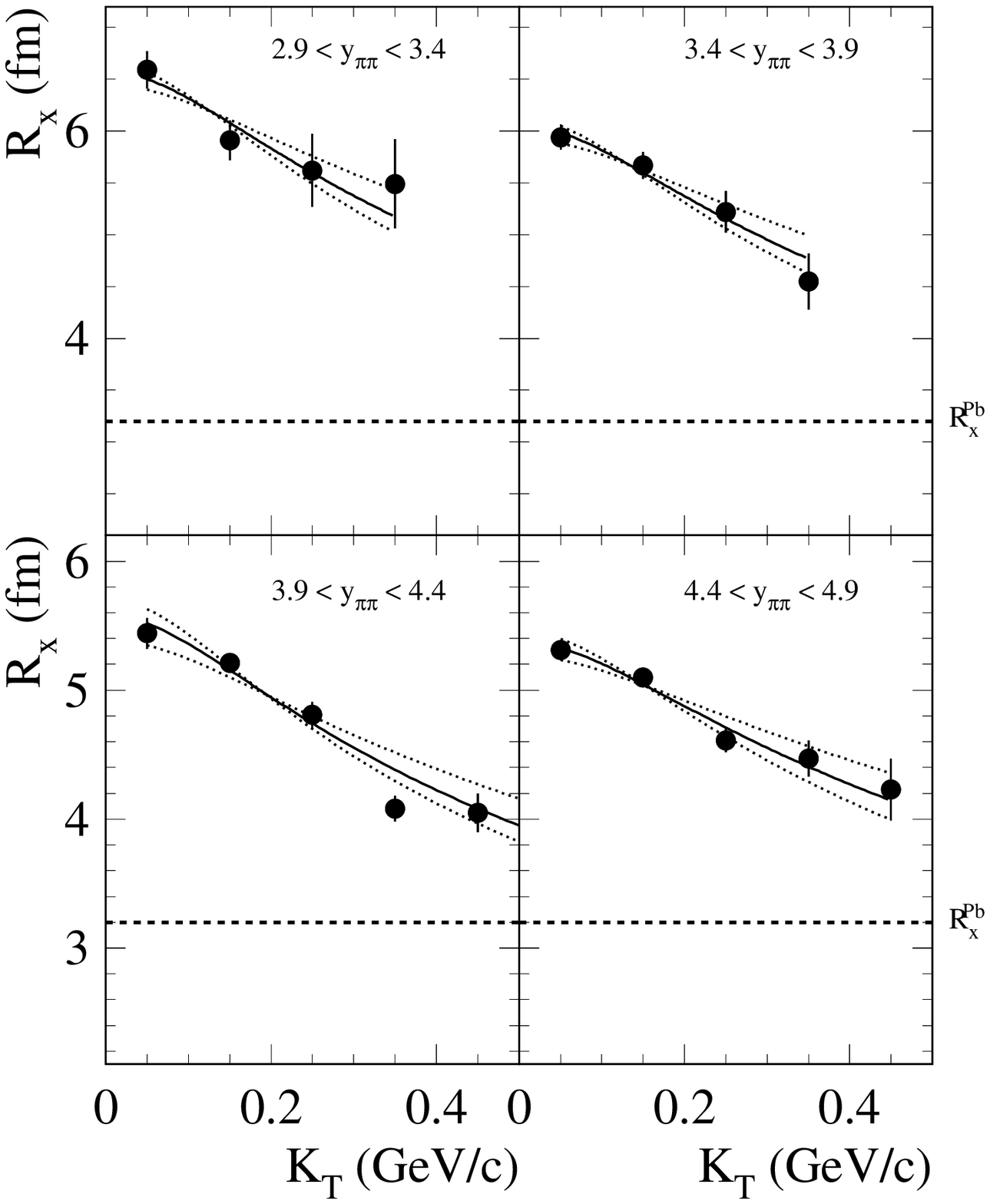,width=1.\linewidth}   
    \end{minipage}
  \end{flushright}
  \begin{flushleft}
  \vspace{-1.5cm}    
\begin{minipage}{0.48\linewidth} 
  \caption{Two-pion correlation functions for central 10.8 A GeV/c Au+Au
collisions at the AGS from E877. The correlation function is shown for the 3
relative momentum components and for positive and negative pions. Figure
from  
\protect\cite{e877_pipi}
\label{figpipi1}
.} 

\end{minipage}
  \end{flushleft}
  \vspace{-5.cm}  
 \begin{flushright}
\begin{minipage}{0.48\linewidth} 
  \caption{Transverse radius parameter as a function of the pion pair
transverse momentum and in different rapidity bins for central Pb+Pb
collisions at the SPS measured by NA49. Also shown is a fit with an
expanding thermal source model. Data and fits from
\protect\cite{na49_pipi}.
The dashed line indicates the measured rms charge radius of Pb divided
by square root of three. 
\label{figpipi2}
}  
\end{minipage}
  \end{flushright}
   \vspace{-0.5cm}  
\end{figure}

Figure~\ref{figpipi1} shows the two-pion correlation function for
positive and negative pions for central Au+Au collisions at the AGS for
the three components of relative momentum together with fits using a
three-dimensional Gaussian parameterization (for details see
\cite{e877_pipi}. Using the radius parameters in the three dimensions to
estimate a lower limit for the volume at freeze-out one obtains V = 2600
fm$^3$ or about twice the volume of the initial Au nucleus; clearly the
system has expanded, in particular in view of the early compression (see
below). The densities at thermal freeze-out for nucleons and pions given
by this volume estimate are 0.11 and 0.12/fm$^3$.

For Pb+Pb collisions at the SPS the two-pion correlation function has
been measured differentially as a function of the pair transverse
momentum. In Figure~\ref{figpipi2} the (one-dimensional) radius
parameter R$_{\rm x}$ is shown as a function of this variable for
different rapidity bins. The decrease with increasing pair transverse
momentum is a clear indication of the expansion of the system. The solid
lines are fits assuming a certain expansion scenario \cite{chapman} and
temperature. Requiring at the same time a consistent description of the
particle spectra an expansion velocity of about 50 \% of the
speed-of-light and a 'true' standard deviation of the x-distribution of
8.2 fm are found together with a temperature of 120 MeV from this
analysis \cite{na49_pipi}. Note that this temperature agrees well with
the one extracted \cite{qm97} based only on spectra and a certain
initial condition (see above). Using the measured pion rapidity density
and the transverse size information one determines a pion density of
0.12/fm$^3$ at thermal freeze-out, a value identical to the one found
for the AGS (see above).

The path from chemical to thermal freeze-out at the SPS is indicated in
Fig.~\ref{figphased} by the arrow and it is defined by the final
temperature and a constant entropy per baryon. At the AGS thermal and
chemical freeze-out closely coincide.

\section{MEMORY OF THE INITIAL CONDITION}

\begin{figure}[b]

\vspace{-1.1cm}

\hspace*{2.5cm}
\epsfig{file=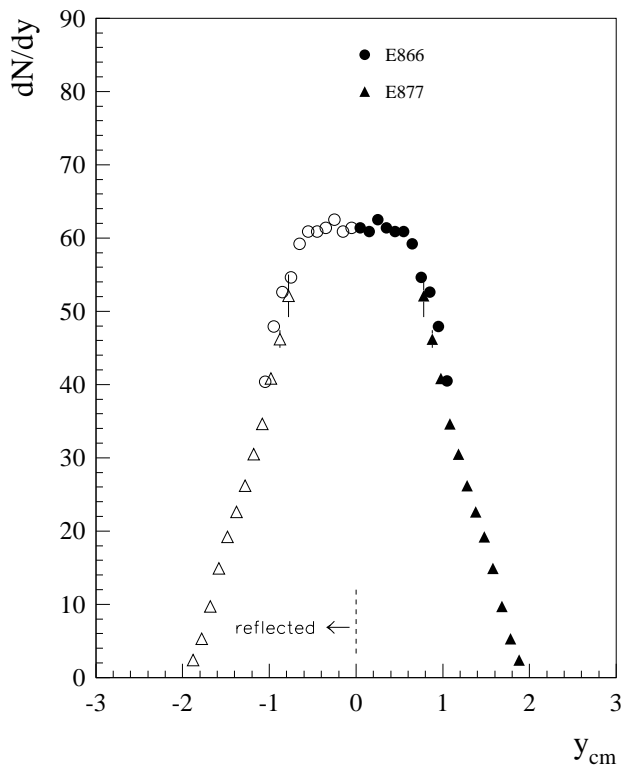,width=10cm}

\vspace{-1.5cm}

\caption{\small Proton rapidity density for central Au+Au collisions at
the AGS at 11 A GeV/c. Data from 
\protect\cite{e877_p,e866_ppi}. 
The open symbols are reflections of the measured points about mid rapidity.
}
\label{figdndy}
\end{figure}

Estimates for the initial baryon and energy density can be obtained from
the rapidity distributions of protons and of transverse energy ${\rm
E_t}$. Figure~\ref{figdndy} shows the distribution of protons in central
Au+Au collisions at the AGS; combining
the data of two experiments the complete distribution is known. The
distribution peaks at midrapidity indicating a very high degree of
stopping. In \cite{qm96} it was shown already that this distribution can
be reproduced quantitatively assuming a uniform distribution of
fireballs over one unit of rapidity in either direction from mid
rapidity. This means the fastest fireballs are receding from the center
with 76 \% of the speed-of-light, or the average longitudinal velocity
is 0.46 c. Since this flat distribution corresponds to a Bjorken-type
boost-invariant scenario (at least over a limited rapidity range) this
same picture can be employed to estimate the early conditions, say at a
time of 1 fm/c or 3 $\cdot$ 10$^{-24}$ s. One finds a nucleon density of
1.1/fm$^3$ - one order of magnitude above the density at thermal
freeze-out - and an energy density of 1.4 GeV/fm$^3$. The same type of
analysis for SPS energies and Pb+Pb collisions gives a baryon density of
0.65/fm$^3$ and an energy density of about 3 GeV/fm$^3$. At higher beam
energy evidently the maximum baryon density is less and the maximum
energy density is higher. Comparing to critical energy densities in
state-of-the-art lattice QCD calculations the energy density at a
the critical temperature is about 1 GeV/cm$^3$. So based on the fact that
the hadrochemical freeze-out points are on or close to the expected
phase boundary and the estimate of the initially significantly higher
energy and baryon densities it is likely that the system has indeed been
beyond the phase boundary to the QGP at an early time. 

\begin{figure}

\vspace{-1.6cm}

\hspace*{1.cm}
\epsfig{file=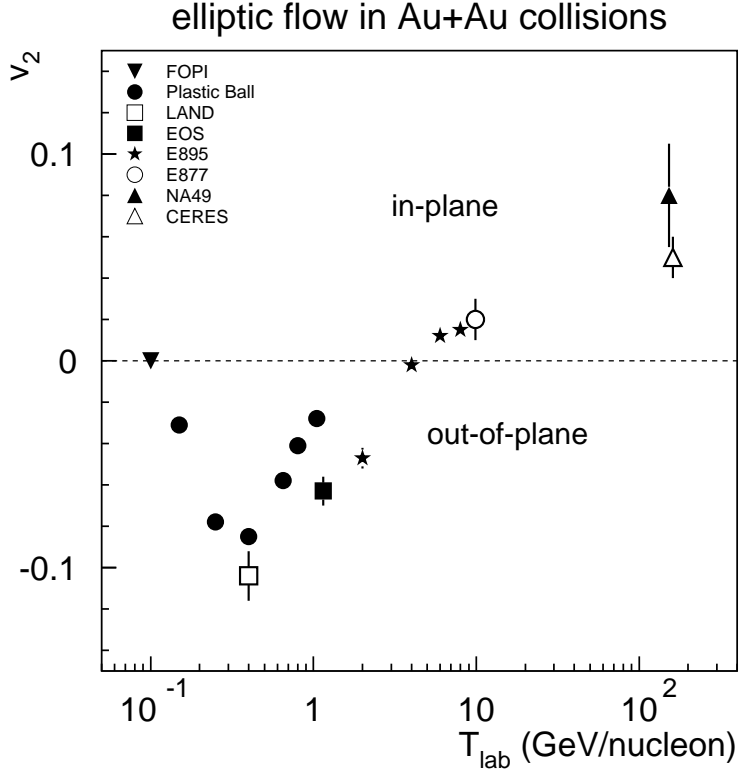,width=13cm}

\vspace{-2.5cm}

\caption{\small Dependence of the elliptic flow parameter v$_2$,
evaluated for nucleons near mid rapidity and for semi-central
collisions, on the beam kinetic energy per nucleon. Data from
\protect\cite{fopi_v2,pb_v2,la_v2,eos_v2,e877_v2,na49_v2,na45_v2}.
Positive values mean preferential emission in the reaction plane. 
}
\label{figv2}
\end{figure}

\enlargethispage{\baselineskip}
Azimuthal distributions of transverse energy and of particle numbers,
when analyzed relative to the reaction plane spanned by the impact
parameter and the beam direction, show clear anisotropies for not too
central collisions. There is a correlation between the final
distributions at freeze-out and the impact parameter. Both a dipole
moment, also called sideways flow, and a quadrupole moment, called
elliptic flow have been found at the AGS and SPS for the heavy colliding
systems. This is another link to the early conditions. I will pick here
only one example and discuss the systematic dependence on beam energy of
the elliptic flow, as it emerges from combining data from a number of
experiments. The elliptic flow is quantified by the second Fourier
coefficient of the azimuthal distribution v$_2$ and the variation with
beam energy is shown in Figure~\ref{figv2} for nucleons close to the
c.m. rapidity.

A negative Fourier coefficient v$_2$ means that the long axis of the
ellipse is perpendicular to the reaction plane (out-of-plane elliptic
flow) while for a positive coefficient it is in the plane (in-plane
elliptic flow). The out-of-plane elliptic flow has been interpreted in
terms of reinteractions of nucleons with projectile/target spectator
nucleons. Naturally, as beam energy increases such remnants will
eventually be too far away from mid-rapidity to affect the distribution
of nucleons there and therefore one had expected this effect to decrease
with increasing beam energy. In-plane elliptic flow develops due to the
asymmetric shape of the overlap volume of the two nuclei and the related
early pressure gradient. It has been predicted \cite{sorge} that the
elliptic flow is sensitive to the early pressure in the collision. 

Even more recently, the beam energy dependence of the elliptic flow has
been studied theoretically for different scenarios, including various
types of equation of state and the possible influence of a phase
transition \cite{daniele}. A phase transition leads to an effective
softening of the equation of state and therefore an upwardly steepening and then flattening S-curve with increasing beam energy. The very recent data from the E895
collaboration map the region between 2 and 8 GeV and there is a close by
point at 10 GeV in good agreement (see Fig.~\ref{figv2}). These data
could just reflect the predicted \cite{daniele} curvature due to a
second order phase transition. The very rich data on flow, also of other
particle species and as a function of transverse momentum and rapidity,
deserve in this light full theoretical exploration, still missing at
present.

\section{VECTOR MESONS IN MEDIUM}

The decay of vector mesons into lepton pairs provides a means to look
into the early, possibly even prehadronic, phase of the heavy ion
collision since final state interactions of the leptons are
negligible. First the low mass region with the $\rho, \omega, \phi$
mesons will be addressed.

The masses of vector mesons as a function of baryon density and
temperature have been discussed in terms of their relation to the
expected restoration of chiral symmetry. A measure for chiral symmetry
breaking is the quark condensate which is found to drop to small values
near the critical temperature in lattice QCD calculations
\cite{lattice}. How this restoration of chiral symmetry is reflected in
observables has been a subject of much discussion (see
e.g. \cite{wambach}). QCD sum rules relate the quark condensate and the
masses of vector mesons such as the $\rho$ or axial vectors such as the
a$_1$, as first applied by \cite{hatsuda}. This idea has led to the
proposal of a linear downward scaling of the vector meson masses with
increasing density, the so-called Brown-Rho scaling
\cite{brownrho}. Later is was argued that the chiral partners such as
the $\rho$ and a$_1$ have to join but that this could happen at a
nonzero mass determined by the value of the vector correlator in the
hadronic medium (see discussion in \cite{wambach}). A whole spectrum of
theoretical possibilities has arisen for the masses and widths of vector
mesons at finite density and temperature (see e.g. Table 3 in
\cite{lee}). Clearly the situation calls for input from measurements.

\begin{figure}[b]
  \begin{flushleft}
\vspace{-1.5cm}    
    \begin{minipage}{0.5\linewidth}
      \epsfig{figure=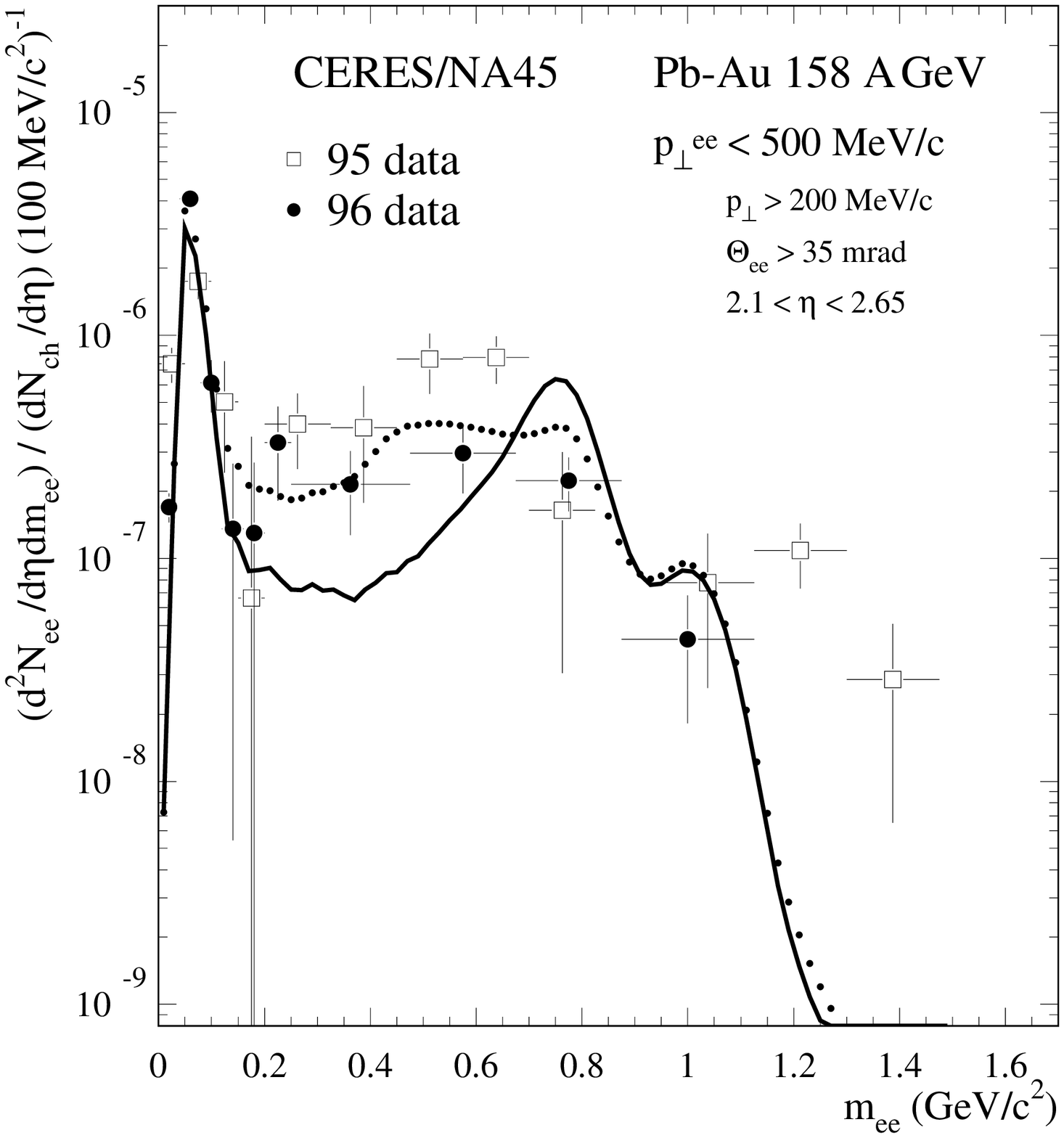,width=1.\linewidth}
    \end{minipage}
  \end{flushleft}
  \vspace{-9.8cm}
  \begin{flushright}
    \begin{minipage}{0.5\linewidth}
      \epsfig{figure=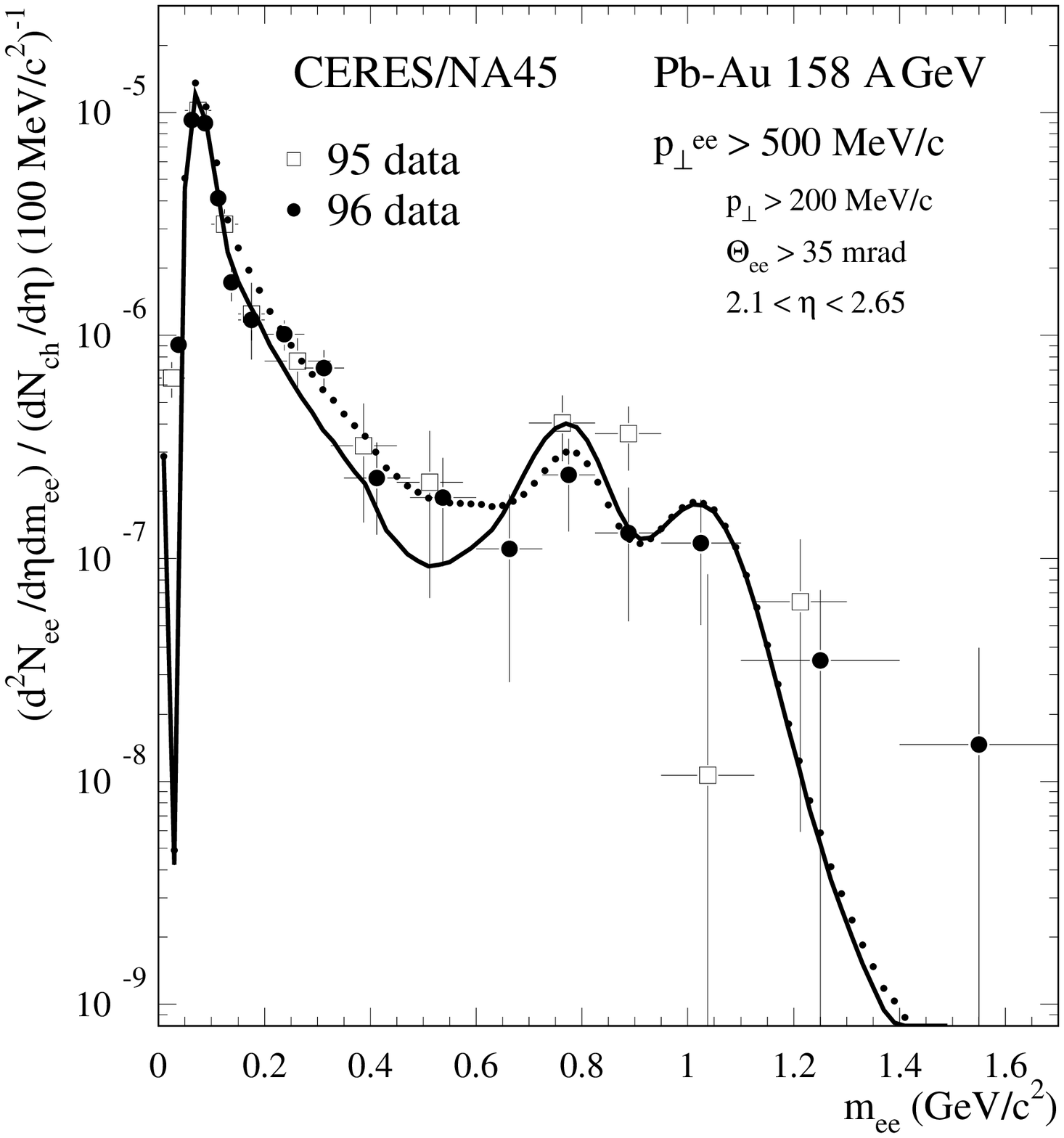,width=1.\linewidth}
    \end{minipage}
  \end{flushright}
\vskip -1.5cm  
\begin{minipage}{1.\linewidth}
  \begin{flushleft}
  \caption{Invariant mass distributions of electron pairs for the most
central third of the geometric cross section in Pb+Pb collisions at 158
A GeV/c obtained by the CERES collaboration from the $'$95 
  and $'$96 data analyses, se\-parated into samples with high and low
pair p$_t$
\protect\cite{lenkeit}.
  The data are compared to model calculations
\protect\cite{rapp}
based on the free $\rho$ properties (solid lines) and employing in-medium
modifications of the $\rho$ (dotted lines).
\label{figee}
}
  \end{flushleft}    

\end{minipage}
   \vspace{-0.7cm}
\end{figure}

An interesting opportunity stems from the short lifetime of the $\rho$
meson. Since it decays on the timescale of one or two fm/c which is
short as compared to the lifetime of the fireball it's decay samples the
hot and dense medium and it's evolution as a function of
time. Experimentally an enhancement as compared to expectations from p+A collisions  has been seen by the CERES
collaboration in the dielectron spectrum in the mass region in between
twice the pion mass and the $\rho, \omega$ pole for both S+Au and Pb+Au
collisions \cite{ceres1,ceres2} and for S+W this is confirmed by the
dimuon measurement from the HELIOS collaboration \cite{helios}. In
\cite{lenkeit} a high statistics measurement for Pb+Au from CERES is
added and the enhancement persists. As compared to simple expectations
for hadron decay contributions the enhancement in the mass region 0.25 -
0.70 GeV/c$^2$ is \cite{lenkeit} 2.6 $\pm$ 0.5(stat.) $\pm$
0.5(syst.). The enhancement is most significant at low pair transverse
momenta and therefore Figure~\ref{figee} shows the pair mass spectrum in
two transverse momentum bins (see also Fig. 3 in \cite{lenkeit}).

It has been argued that in the dense hadronic fireball processes such as
pion annihilation could be an important source of dilepton pairs. Indeed
this possibility is supported by the approximately quadratic scaling of
the electron pair yield in this mass region \cite{lenkeit}. However,
calculations show that pion annihilation is dominated by the $\rho$
resonance and, while there is indeed a significant enhancement by this
mechanism, the shape of the experimental spectrum is not reproduced at
all. This is exemplified by the calculations shown as the solid lines in
Fig.~\ref{figee}. Introducing in-medium modifications of the $\rho$
meson changes the situation drastically. One approach, coupling the
$\rho$ meson to baryons \cite{rw}, leads to a significant downward shift
of the $\rho$ pole and a drastic increase in the width. This is the
basis of the calculations shown as the dotted lines in Fig.~\ref{figee}
and it can be seen that now the experimental data can be accounted
for. The effect of the Brown-Rho scaling is similar \cite{cracow,rapp}
although in detail there are differences. In order to differentiate
between different possibilities and to also study the fate of the
$\omega$ meson for which very interesting effects have been predicted as
well \cite{klingl} data with higher resolution and still higher
statistics are required.

While the modifications of the $\rho$ and $\omega$ mesons address the
state of the fireball between beginning hadronization and thermal
freeze-out a much earlier time-scale is probed by the J/$\Psi$
meson. The charm quark pairs are thought to be created in hard early
collisions, in the kinematic regime relevant in present heavy ion
experiments in gluon fusion. The question whether they hadronize into a
J/$\Psi$ or $\Psi$' meson is determined by the medium surrounding
them. This study was fueled by the longstanding prediction by Satz and
Matsui \cite{sm} that the heavy quark effective potential is screened by
a high density of partons and thereby the J/$\Psi$ and similar mesons
melt in a process analogue to the Debye screening.

\begin{figure}[t]

\vspace{-0.5cm}

\hspace*{1.cm}
\epsfig{file=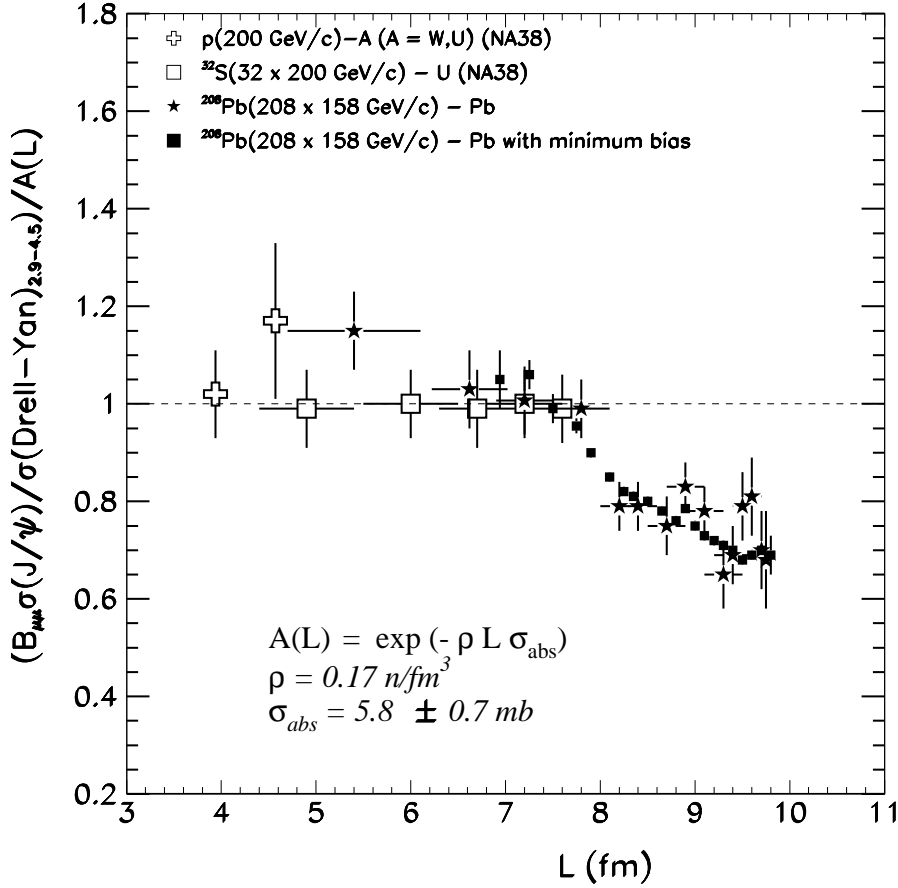,width=12cm}

\vspace{-.9cm}

\caption{\small J/$\Psi$ production normalized to the Drell-Yan yield as
a function of the length L of nuclear matter traversed by the J/$\Psi$.
Figure from
\protect\cite{na50}.
}
\label{figna50}
\end{figure}



The production of J/$\Psi$ mesons was studied from the beginning of the
CERN heavy ion program. A suppression relative to the continuum was seen
in O+Cu,U and S+U collisions and this was complemented by a systematic
study of p+nucleus collisions (a summary can be found in
\cite{lourenco}). The systematics of J/$\Psi$ production in p+A and
O,S+A data can be summarized by a power law dependence of the J/$\Psi$
production cross section as (A$\cdot$B)$^\alpha$, where A and B are the
mass numbers of target and projectile and $\alpha$ was determined as
0.911$\pm$0.016.

In order to include measurements at various centralities for heavy ion
collisions into the systematics a new variable L was introduced which
measures the path length of nuclear matter traversed by the J/$\Psi$
assuming a straight trajectory. The production cross section normalized
to A$\cdot$B was found to fall exponentially \cite{gh} as A(L) =
exp(-$\rho L \sigma{\rm_{abs}}$) with $\sigma{\rm_{abs}}$ of 6-7 mb and
the nuclear matter density $\rho$ = 0.17/fm$^3$. This surprizingly large
cross section was attributed to the absorption of a color singlet
(c\=cg) state that is the precursor of the fully formed hadron
\cite{ks}. The data for Pb+Pb collisions show a significant deviation
from the scaling behavior just discussed. Figure~\ref{figna50} displays
the cross section for J/$\Psi$ production multiplied with the branching
ratio for the dimuon decay and normalized to the Drell-Yan cross section
divided by the exponential fall-off A(L).






The Pb+Pb results are displayed for two different analyses as a function
of centrality and for both one can see that the results for more
peripheral collisions (smaller L) are close to the systematics of
exponential fall-off obtained for lighter systems while the points for
more central collisions (impact parameters below about 8 fm) exhibit an
additional suppression. The stars represent data that are normalized to
the Drell-Yan cross section measured in the same experiment and the
statistical errors are dominated by the much smaller Drell-Yan cross
section. The squares avoid this statistical uncertainly by instead
utilizing a measured minimum bias transverse energy distribution and a
calculated ratio of Drell-Yan to minimum bias cross section for every
centrality (or E$_{\rm t}$). Again a systematic drop is visible for b
below 8 fm (L above 8 fm). Comparison of the two analyses establishes
that the drop can only be due to a reduced J/$\Psi$ cross section.

Since the drop in J/$\Psi$ production appears at a certain centrality in
Pb+Pb collisions it is indeed plausible that this is related to a
certain critical energy density, possibly reached over a certain
critical volume. It has been demonstrated recently \cite{sriva} in a
parton cascade 
calculations that at SPS energy and a time of 0.3 fm/c the energy
density in the center of the colliding nuclei exceeds 9 GeV/fm$^3$ while
in S+S collisions a value of 3 GeV/fm$^3$ is never exceeded. This idea
has been exploited in a model 
\cite{khar,khar2} where formation of QGP bubbles is assumed and above
a critical density of 1.4 GeV/fm$^3$ and a radius of 4 fm all $\chi_c$
are dissociated; the 40 \% portion of J/$\Psi$ that is produced via
decays of the $\chi_c$ thus disappears. At higher energy densities also
the J/$\Psi$ itself is expected to  melt in the plasma. This model still
requires some ad hoc assumptions (see above) and any alternative
solution needs to be excluded. On the other hand there is now a body of
high quality data for different systems and centralities including the
transverse momentum distribution of the J/$\Psi$ mesons \cite{topil} and
any theoretical interpretation needs to address in a consistent way all
available data.

\section{OUTLOOK}

Present fixed target experiments at the AGS and SPS studying collisions
between the heaviest nuclei exhibit a number of interesting features. The
hadron yields are frozen at or very near the expected phase boundary to
the QGP. At this point strangeness is, contrary to p+p collisions, also in
equilibrium, i.e. saturated, even for rarely produced multiply strange
hadrons. It is not clear how this equilibration, also of particle species
which are expected to have small production cross sections, can come about in a
purely hadronic scenario and on the time scales available in this
case. This, together with the circumstantial evidence
that the initial baryon and energy density at AGS and SPS are beyond the
estimated critical values early in the collision, makes it likely that
indeed the system is in the QGP phase for some time before it hadronizes
again. Estimates of the maximum temperature reached give values of about
200 MeV, i.e. not much beyond the critical temperature. This could still
result in a visible signal in direct photons radiated during the
evolution if the hadronization takes long enough. There is tantalizing
first indication that there may indeed be a signal \cite{photons}.
The anisotropic flow and in particular the elliptic flow and their beam
energy dependence are sensitive to the equation of state and changes in
it. The values around the zero crossing at the AGS could be related to a
softening of the equation of state associated with a transition from
hard hadronic matter to softer quark matter. The enhancement in the
dilepton mass spectrum above twice the $\pi$ and below the $\rho$ mass
indicates in-medium
modifications of the $\rho$ meson. Their relation to chiral symmetry
restoration and possible modifications of the other light vector mesons
need more experimental and theoretical study. The unexpected additional
suppression of the J/$\Psi$ for more central Pb+Pb collisions as
compared to previously studied p+p, p+A, and A+B systems could indicate
the predicted color screening in the QGP. But a consistent
implementation of other possible scenarios and a comparison to all
availalbe data is still needed to see whether this is the only
explanation of this intriguing signal.


\begin{figure}[h]

\vspace{-2.cm}

\hspace*{1.cm}
\epsfig{file=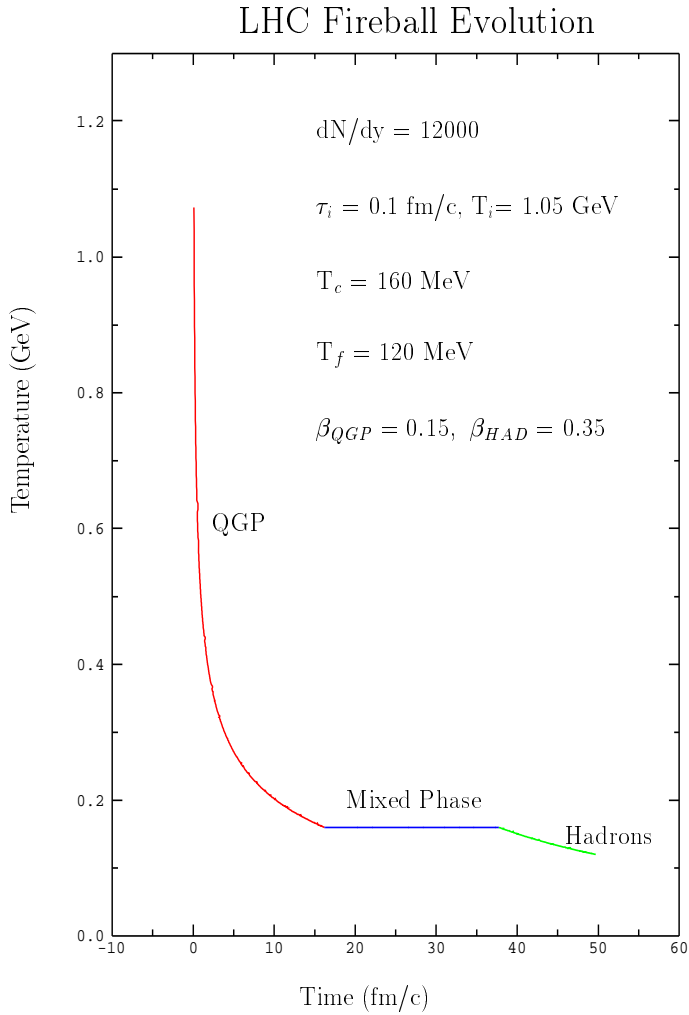,width=12cm}

\vspace{-3.cm}

\caption{\small Expected evolution of the fireball after a Pb+Pb
collision at LHC. Figure from
\protect\cite{pbm}.
}
\label{figlhc}
\end{figure}

The advent of the colliders RHIC at Brookhaven later this year and of
the LHC in 2005 will add a new dimension to the study of the QGP. At
these accelerators collisions of the heaviest nuclei will be studied at
c.m.  energies a factor of 12 and 330 higher than presently available at
the SPS. Based on the new information on parton distributions in the
nucleon from HERA this leads to a large number of partons in the early
phase and to a rapid thermal equilibration. It is estimated that there
will be about 4000 gluons per unit rapidity above p$_t$ = 2 GeV/c and
about a factor of ten less quarks leading to very copious (mini)jet
formation. This implies that  the system equilibrates thermally on the
timescale of 0.1 fm/c and would lead to initial temperatures of 450
and 1000 MeV at RHIC and LHC, respectively. Both values are well beyond
the critical temperature and should allow to study the (ideal) parton
gas and it's evolution and eventual hadronization.

A typical expectation for the evolution of the system after the initial
short equilibration phase is shown in Figure~\ref{figlhc} for a Pb+Pb
collision at the LHC. With the somewhat conservative expectation that
transverse expansion, and therefore more rapid cooling, is already
significant in the plasma phase the plasma expands and cools to the
critical temperature over about 10-15 fm/c \cite{pbm}. Then
hadronization takes place and since a large number of degrees of freedom
has to be converted into a smaller number corresponding to a hot hadron
gas the volume has to grow accordingly (so that entropy stays at least
constant). This takes typically 20 fm/c. The final expansion of the
hadronized system until thermal freeze-out is presumably similar to what
is currently already seen at the SPS and this final point could again be
determined by the pion density of 0.1/fm$^3$. With the expected very
large pion multiplicities (about 8000 per unit rapidity at LHC) this
leads also to a very huge volume in the final state.

With the very large initial temperatures and the clearly partonic nature
of the early state new signatures and tools become available. A
comprehensive discussion is beyond the scope of the present discussion.
Among the interesting features to be studied are the energy loss of
partons in a QGP, the associated question of jet quenching, direct
thermal radiation of the very hot plasma via q\=q annihilation and the
QCD Compton process, color screening of J/$\Psi$, $\Psi$', and
$\Upsilon$ states, open charm and beauty production, in addition to the
array of phenomena studied already presently at the lower energies.


\begin{thebibliography}{20}

{
\bibitem{lattice} See e.g. the proceedings of the Lattice'98 and
preceeding conferences, Nucl. Phys. {\bf B63} (1998).

\bibitem{lattice1} C. Bernard et al., MILC collaboration,
Phys. Rev. {\bf D56} (1997) 5584.

\bibitem{lattice2} E. Laermann, Nucl. Phys. {\bf B63} (1998) 114.

\bibitem{qm96} J. Stachel, Nucl. Phys. {\bf A610} (1996) 509c.

\bibitem{thermal1} P. Braun-Munzinger, J. Stachel, J.P. Wessels, N. Xu,
Phys. Lett. {\bf B344} (1994) 43.

\bibitem{wa97} E. Andersen et al., WA97 collaboration, preprint
CERN-EP/99-29, Phys. Lett. {\bf B} in print.

\bibitem{therm_pbpb} P. Braun-Munzinger, I. Heppe, J. Stachel, preprint
nucl-th/9903010 and subm. Phys. Lett. {\bf B}.

\bibitem{becattini} F. Becattini, J. Phys. {\bf G23} (1997) 1933.

\bibitem{phase_gerry} P. Braun-Munzinger and J. Stachel,
Nucl. Phys. {\bf A606} (1996) 320.

\bibitem{qm97} P. Braun-Munzinger and J. Stachel, Nucl. Phys. {\bf A638}
(1998) 3c.

\bibitem{averbeck} Modifications to Fig.
\protect\ref{figphased}
courtesy of Ralf Averbeck.

\bibitem{laer} E. Laermann, Nucl. Phys. {\bf A610} (1996) 1c.

\bibitem{cley} J. Cleymans and K. Redlich, GSI preprint-98-43.

\bibitem{na44sl} I.G. Bearden et al., NA44 collaboration,
Phys. Rev. Lett. {\bf 78} (1997) 2080; M. Kaneta for the NA44
collaboration, Nucl. Phys. {\bf A638} (1998) 419c; A. Sakaguchi for the
NA44 collaboration, ibidum 103c.

\bibitem{na49sl} G. Roland for the NA49 collaboration, Nucl. Phys. {\bf
A638} (1998) 91c; F. P\"uhlhofer for the NA49 collaboration, ibidum
431c; H. Appelsh\"auser et al., NA49 collaboration, Phys. Lett. {\bf
B444} (1998) 523.

\bibitem{wa97sl} E. Andersen et al., WA97 collaboration,
Phys. Lett. {\bf B433} (1998) 209; R. Lietava for the WA97 collaboration
in Proc. Strangeness 98, J. Phys. G: Nucl. Part. Phys. {\bf 25} (1999).

\bibitem{wa98sl} M.M. Aggarwal et al., WA98 collaboration,
Nucl. Phys. {\bf A638} (1998) 147c.

\bibitem{sr} One of the earliest references where this connection is made
is P.J. Siemens and J.O. Rasmussen, Phys. Rev. Lett. {\bf 42} (1979) 880.

\bibitem{thermal2} P. Braun-Munzinger, J. Stachel, J.P. Wessels, N. Xu,
Phys. Lett. {\bf B365} (1996) 1.

\bibitem{esumi} S. Esumi, S. Chapman, H. van Hecke, N. Xu,
Phys. Rev. {\bf C55} (1997) R1. 

\bibitem{e877_pipi} J. Barrette et al., E877 collaboration,
Phys. Rev. Lett. {\bf 78} (1997) 2916.

\bibitem{na49_pipi} H. Appelsh\"auser et al., NA49 collaboration,
Eur. Phys. J. {\bf C2} (1998) 661.

\bibitem{chapman} S. Chapman, J.R. Nix, U. Heinz, Phys. Rev. {\bf C52}
(1995) 2694.

\bibitem{e877_p} R. Lacasse, Ph.D. thesis McGill University, 1998;
J. Barrette et al., E877 collaboration, manuscript in preparation.

\bibitem{e866_ppi} L. Ahle et al., E866 collaboration, Phys. Rev. {\bf C57}
(1998) R466.

\bibitem{fopi_v2} Bastid et al., FOPI collaboration, Nucl. Phys. {\bf
A622} (1997) 573.

\bibitem{pb_v2} H.H. Gutbrod, K.H. Kampert, B. Kolb, A.M. Poskanzer,
H.G. Ritter, R. Schicker, H.R. Schmidt, Phys. Rev. {\bf C42} (1990) 640.

\bibitem{la_v2} Y. Leifels et al., LAND collaboration,
Phys. Rev. Lett. {\bf 71} (1993) 963.

\bibitem{eos_v2} C. Pinkenburg et al., E895 collaboration,
subm. Phys. Rev. Lett. (1999).

\bibitem{e877_v2} J. Barrette et al., E877 collaboration,
Phys. Rev. {\bf C55} (1997) 1420.

\bibitem{na49_v2} H. Appelsh\"auser et al., NA49 collaboration,
Phys. Rev. Lett. {\bf 80} (1998) 4136.

\bibitem{na45_v2} F. Ceretto for the CERES collaboration,
Nucl. Phys. {\bf A638} (1998) 467c.

\bibitem{sorge} H. Sorge, Phys. Rev. Lett. {\bf 78} (1997) 2309.

\bibitem{daniele} P. Danielewicz et al., Phys. Rev. Lett. {\bf 81}
(1998) 2438.

\bibitem{wambach} J. Wambach and R. Rapp, Nucl. Phys. {\bf A638} (1998)
171c. 

\bibitem{hatsuda} T. Hatsuda and S.H. Lee, Phys. Rev. {\bf C52} (1992)
R34.  

\bibitem{brownrho} G.E. Brown and M. Rho, Phys. Rev. Lett. {\bf 66}
(1991) 2720.

\bibitem{lee} S.H. Lee, Nucl. Phys. {\bf A638} (1998) 183c.

\bibitem{ceres1} G. Agakichiev et al., CERES collaboration,
Phys. Rev. Lett. {\bf 75} (1995) 1272.

\bibitem{ceres2} G. Agakichiev et al., CERES collaboration, Phys. Lett. {\bf
B422} (1998) 405.

\bibitem{helios} M. Masera for the HELIOS collaboration,
Nucl. Phys. {\bf A590} (1995) 93c.
 
\bibitem{lenkeit} B. Lenkeit for the CERES collaboration, these
proceedings.

\bibitem{rapp} R. Rapp, proceedings Rencontres de Moriond '98,
nucl-th/9804065; R. Rapp and C. Gale, hep-ph/9902268; and R. Rapp,
private communication.

\bibitem{rw} R. Rapp, G. Chanfray, J. Wambach, Nucl. Phys. {\bf A617}
(1997) 472.

\bibitem{cracow} G.E. Brown, G.Q. Li, R. Rapp, M. Rho, J. Wambach, Acta
Phys. Polon. {\bf B29} (1998) 2309. 

\bibitem{klingl} F. Klingl, N. Kaiser, W. Weise, Z. Phys. {\bf A356}
(1996) 193; Nucl. Phys. {\bf A624} (1997) 527.

\bibitem{sm} T. Matsui and H. Satz, Phys. Lett. {\bf B178} (1986) 416.

\bibitem{lourenco} C. Lourenco, Nucl. Phys. {\bf A610} (1996) 552c.

\bibitem{gh} C. Gerschel and J. H\"ufner, Nucl. Phys. {\bf A544} (1992)
513c.

\bibitem{ks} D. Kharzeev and H. Satz, Phys. Lett {\bf B366} (1996) 316.

\bibitem{na50} M.C. Abreu et al., NA50 collaboration, preprint
CERN-EP/99-13 and Phys. Lett. {\bf B} (1999) in print.

\bibitem{sriva} D.K. Srivastava and K. Geiger, Nucl. Phys. {\bf A647}
(1999) 136.

\bibitem{khar} D. Kharzeev, Nucl. Phys. {\bf A638} (1998) 279c.

\bibitem{khar2} D. Kharzeev, M. Nardi, H. Satz, preprint hep-ph/9707308.

\bibitem{topil} N.S. Topilskaya for the NA50 collaboration, these
proceedings.
\bibitem{photons} V. Manko for the WA98 collaboration, these proceedings.
\bibitem{pbm} P. Braun-Munzinger, private communication.

}
\end{thebibliography}
\end{document}